\documentstyle[preprint,aps,floats]{revtex}
\begin{document}
\preprint{\tighten \vbox{\hbox{NSF-ITP-00-47}
  \hbox{hep-th/0006013}}}
\draft
\title{Comments on Noncommutative Open 
String Theory:\\
V-duality and Holography}
\author{Guang-Hong Chen$^{1,2}$, 
and Yong-Shi Wu$^{1}$}
\address{1. Department of Physics, University 
of Utah\\ Salt Lake City, Utah  84112\\
          2. Institute for Theoretical Physics\\ 
University of California, Santa Barbara, CA 93106 \\
\vspace{.5cm} 
{\tt ghchen@physics.utah.edu \\
     wu@physics.utah.edu}
}

\maketitle

{\tighten
\begin{abstract}
In this paper we study the interplay of 
electric and magnetic backgrounds in 
determining the decoupling limit of coincident 
D-branes towards a noncommutative Yang-Mills 
(NCYM) or open string (NCOS) theory. No decoupling
limit has been found for NCYM with space-time 
noncommutativity. 
It is suggested that there is a new duality, 
which we call {\it V-duality}, which acts on 
NCOS with both space-space and space-time 
noncommutativity, resulting from decoupling 
in Lorentz-boost related backgrounds. We also 
show that the holographic correspondence,
previously suggested by Li and Wu, between 
NCYM and its supergravity dual can be 
generalized to NCOS as well. 
\end{abstract}
}
\newpage


\section{Introduction}

A noncommutative space or spacetime is
the one with noncommuting coordinates,
satisfying 
\begin{equation}
[x^\mu, x^\nu]=i\theta^{\mu\nu} \hspace{1cm} 
\mu, \nu =0,1,2,\cdots, 
\label{noncom}
\end{equation}
where $\theta^{\mu\nu}$ are antisymmetric 
and real parameters of dimension length
squared. A field theory on such a space
can be formulated using a representation, 
in which the coordinates $x^\mu$ are the 
same as usual, but the product of any two 
fields of $x^\mu$ is deformed to the 
Moyal star-product:
\begin{equation} 
\label{starproduct}
f*g(x)=e^{(i/2)\theta^{\mu\nu}
\partial^{x}_{\mu}
\partial^{y}_{\nu}} f(x)g(y)|_{y=x},
\label{star}
\end{equation}
while the commutator in Eq. (\ref{noncom}) 
is understood as the Moyal bracket with 
respect to the star product:
\begin{equation}
[x^{\mu},x^{\nu}]\equiv x^{\mu}*x^{\nu}
-x^{\nu}*x^{\mu}.
\end{equation}

Recently it has been shown that Yang-Mills 
theory (or open string theory) on such 
noncommutative space (or spacetime), which 
we will abbreviate as NCYM (or NCOS), arises 
naturally in string or M(atrix) theory on 
coincident D-brane world-volume in 
anti-symmetric tensor backgrounds in certain 
scaling limits (decoupling or DLCQ limits) 
\cite{matrix,openstring,sei-witten,seiberg,gop}.
(In order to obtain a nontrivial theory defined 
only on the brane world-volume, these scaling 
limits require that in addition to the usual 
$\alpha^{'}\rightarrow 0$ limit, certain components of 
the closed-string metric and/or those of the 
background parallel to the brane world-volume 
should also be scaled in appropriate way. For
details, see refs. \cite{sei-witten,seiberg,gop}.)
This strongly suggests that space-space or 
even space-time noncommutativity could be 
a general feature of the unified theory of 
quantum gravity at a generic point inside 
the moduli space of string/M theory. 
Though perhaps not every noncommutative field 
(or string) theory is a consistent quantum 
theory on its own, there is a belief that 
noncommutative field (or string) theories
that can arise as effective limits in 
fundamental string theory should be 
consistent quantum theory on their own.
Up to now, only NCYM with space-space 
noncommutativity and NCOS with space-time 
noncommutativity have been obtained by
taking certain decoupling limits in string 
theory. It is important to clarify whether 
there exist decoupling limits in string theory 
with backgrounds that lead to either NCYM or 
NCOS with both space-space and space-time 
noncommutativity. Namely one wants to know 
how big the moduli space is for NCYM and NCOS 
that can arise from string theory.

Constant bulk B-background in string theory 
in topologically trivial spacetime can be 
gauged away, while inducing 
constant gauge field background on the 
D-brane world-volume. One might wonder 
whether the nature, electric or magnetic, 
of the gauge background would affect the 
scaling limit that decouples the theory
on the D-branes from closed strings in 
the surrounding bulk. From the 
Born-Infeld action it is known that on 
the D-brane no electric field can be 
stronger than a critical electric field, 
while the same is not true for magnetic 
fields. Indeed recent careful 
reexamination of the decoupling limits 
shows that though the decoupling limit 
in a magnetic background always results 
in NCYM with only space-space 
noncommutativity \cite{sei-witten}, 
in an electric background the decoupling 
limit becomes different and leads to 
NCOS with only space-time noncommutativity
 \cite{seiberg,gop,barbon}. 
Moreover, theory with space-time 
noncommutativity is expected to behave 
very differently from one with only
space-space noncommutativity. Recently
whether a field theory with space-time 
noncommutativity is unitary has been 
questioned in the literature 
 \cite{toum,gomis}. All these inspire
the following questions: What would 
happen if there are both electric and 
magnetic backgrounds? Could an NCYM 
with space-time noncommutativity, or 
an NCOS with both space-space and 
space-time noncommutativity, arises 
in favorable situations? 

The present paper will address the 
problem of the interplay of constant 
electric and magnetic backgrounds 
in determining the decoupling limit 
towards a noncommutative theory on 
the D-brane world-volume. To simplify, 
we will restrict ourselves to the 
case of the D3-brane(s). Generalizing 
to other D$p$-branes should be 
straightforward. We will consider 
two special  cases, in which the 
electric and magnetic backgrounds 
are either {\it parallel} or 
{\it perpendicular} to each other. 
It is known that the endpoints of 
an open string behave like (opposite) 
charges on the D3-brane, and the 
motion of a charge in the above two 
background configurations is very 
different. So we expect that there 
should be important differences 
between the decoupling limits in 
the above two cases. As we will 
show in Sec. 2, in either case 
{\it no} decoupling limit can be 
found to lead to an {\it NCYM with 
space-time} noncommutativity. On 
the other hand, in Sec. 3 we will 
show that in favorable situations 
appropriate decoupling limit may 
result in {\it NCOS with both
space-space and space-time} 
noncommutativity. 

In electrodynamics it is known that Lorentz 
boosts act on constant electromagnetic 
backgrounds. Through the decoupling limit
the latter, in turn, affects the 
noncommutativity parameters that define
the resulting NCOS. Thus, the NCOS that 
result from Lorentz-boost related 
backgrounds should be equivalent to each 
other, describing the same decoupled D-brane 
system. We will call this exact equivalence 
among NCOS as {\it V-duality}, which can be 
viewed as the fingerprint of the antecedent 
Lorentz-boost action surviving the decoupling 
limit.  In Sec. 4 we will identify some 
orbits of V-duality in the moduli space of 
NCOS (with both space-space and space-time 
noncommutativity). 

Previously Li and one of us 
 \cite{LiWu} have shown that there 
is a running holographic correspondence 
between NCYM and its gravity dual. Namely, 
the radial dependence of the profile of 
NSNS fields in the gravity dual of an 
NCYM can be derived from the Seiberg-Witten 
relations \cite{sei-witten} between 
close string moduli and open string moduli,
provided that the string tension is running 
with a simply prescribed dependence on the 
energy scale, which is identified with the 
radial coordinate by the well-known UV-IR 
relation \cite{suss-witten}. We 
will show in Sec. 5 that the Li-Wu 
holography argument can be generalized to
NCOS, though with a different prescription 
for the running string tension. 

\section{Decoupling limit for NCYM}

In this section, we concentrate on the 
decoupling limit for NCYM, when the gauge 
background $B_{\mu\nu}$ on a flat 
D$p$-brane world volume (with a constant 
metric $g_{\mu\nu}$) has both electric and 
magnetic components. For definiteness, we 
consider the case with $p=3$. To be specific, 
we restrict ourselves to the special cases with 
the electric and magnetic fields are either 
perpendicular or parallel to each other. 
The generalization to the most general 
configuration should be straightforward.

A constant $B$-background on the D-brane 
does not affect the equations of motion for 
open strings, while it changes the open string 
boundary conditions to
\begin{equation}
\label{boundary}
g_{\mu\nu}\partial_{n}X^{\nu}
+2\pi\alpha^{\prime}B_{\mu\nu}\partial_{s} 
X^{\nu}|_{\partial\Sigma}=0,
\end{equation} 
where the operators $\partial_{n}$ and 
$\partial_{s}$ are the derivatives 
normal and tangential to the 
worldsheet boundaries $\partial\Sigma$. 
For the disc topology, the propagator 
along the boundary is known to be 
 \cite{fradkin,callan} 
\begin{equation}
\label{propagator}
<x^{\mu}(\tau)x^{\nu}(0)>=
-\alpha^{\prime}G^{\mu\nu}\ln(\tau^{2})
+i\frac{\theta^{\mu\nu}}{2}
\varepsilon(\tau).
\end{equation}
As emphasized by Seiberg and Witten in 
Ref. \cite{sei-witten}, the 
physics behind these equations is that 
the moduli ($G_{\mu\nu}$, 
$\theta^{\mu\nu}$, $G_s$) seen by 
open string ends on the D-brane are 
very different from those ($g_{\mu\nu}$, 
$B_{\mu\nu}$, $g_s$) seen by close 
strings; they are related by the  
following elegant relations 
 \cite{sei-witten}
\begin{equation}
\label{Gmunu}
G_{\mu\nu}=g_{\mu\nu}
-(2\pi\alpha^{\prime})^2(Bg^{-1}B)_{\mu\nu},
 \end{equation}
\begin{equation}
\label{inverseGmunu}
G^{\mu\nu}=\biggl
(\frac{1}{g+2\pi\alpha^{\prime}B}
\biggr)_{S}^{\mu\nu},
 \end{equation}
\begin{equation}
\label{theta}
\theta^{\mu\nu}=2\pi\alpha^{\prime}
\biggl(\frac{1}{g+2\pi\alpha^{\prime}B}
\biggr)_{A}^{\mu\nu},
\end{equation}
\begin{equation}
\label{coup}
G_s=g_s\biggl
(\frac{\mbox{det}G_{\mu\nu}}
{\mbox{det}(g_{\mu\nu}
+2\pi\alpha^{\prime}B_{\mu\nu})}
\biggr)^{\frac{1}{2}},
\end{equation}
where $()_S$ and $()_A$ denote, respectively, 
the symmetric and anti-symmetric parts, and 
$G_s$,  $g_s$ the open string and closed 
string coupling.

For the purely magnetic case (with $B_{0i}=0$), 
the scaling limit that decouples the theory on the 
D-brane from closed strings in the bulk has been 
analyzed in Ref. \cite{sei-witten}, and 
may be summarized as a limit subject to the 
following conditions: 
(1) $\alpha^{\prime}\to 0$;
(2) $G_{\mu\nu}$ is finite;
(3) $\theta^{\mu\nu}$ is finite.
The decoupling limit results in an NCYM on 
the D-brane world volume.  For the purely 
electric case (with $B_{ij}=0$), the above 
decoupling limit has been shown 
 \cite{seiberg} not to exist, 
because of the existence of a critical electric 
field-strength. In the following, we will carry 
out an analysis for the case with both electric 
and magnetic components present in the 
antisymmetric tensor $B_{\mu\nu}$. The
interplay between the electric background
(${\bf E}$) and the magnetic background
(${\bf B}$) is worthwhile to explore, since 
in the presence of both an electric field and 
a magnetic field the dynamical behavior of a 
point charge, representing an endpoint of the 
open string, is known to be very different from 
the case in either a purely magnetic or a purely 
electric field. For simplicity, we assume that
either ${\bf E}\perp {\bf B}$ or ${\bf E}
\parallel {\bf B}$. In this section, we discuss
whether a decoupling limit leading to NCYM 
exists in these two cases. 

First, let us consider the case with $B_{01}=E$ 
and $B_{12}=B$, all other components being zero;
namely, the tensor $B_{\mu\nu}$ takes the form
 (for $\mu,\nu=0,1,2$)
\begin{equation}
\label{bper}
 B_{\mu\nu}=\left(
\begin{array}{ccc}
0&E&0 \\
-E&0&B\\
0&-B&0
\end{array}\right).
\end{equation}
The closed string metric $g_{\mu\nu}$ is 
taken to be of the diagonal form
\begin{equation}
\label{metric1}
 g_{\mu\nu}=\left(\begin{array}{ccc}
-g_0&0&0 \\
0&g_1&0\\
0&0&g_2
\end{array}\right).
\end{equation}
For convenience, we follow Ref. 
\cite{seiberg} and \cite{gop} to 
introduce the
critical value, $E_{c}$, of the electric field 
\begin{equation}
\label{critical}
E_{c}=\frac{\sqrt{g_0g_1}}
{2\pi\alpha^{\prime}}.
\end{equation}

Substituting Eq. (\ref{bper}) and Eq. 
(\ref{metric1}) into the Seiberg-Witten 
relations Eqs. (\ref{Gmunu})-(\ref{coup}), 
we get
\begin{equation}
\label{GmunuBperp}
G_{\mu\nu}=\left(\begin{array}{ccc}
-g_0(1-e^2)&0&-g_0eb \\
0&g_1(1-e^2)+\frac{g_0g_1}{g_2}b^2&0\\
-g_0eb&0&g_2+g_0b^2
\end{array}\right)
\end{equation}

\begin{equation}
\label{thetaperp}
\theta^{\mu\nu}=
\frac{1}
{[g_2(1-e^2)+g_0b^2] E_c}
\left(\begin{array}{ccc}
0&g_2e&0 \\
-g_{2}e&0&-g_0b\\
0&g_0 b&0
\end{array}\right),
\end{equation}
\begin{equation}
\label{effectiveGper}
G_s=g_s\sqrt{1+\frac{g_0}{g_2}b^2-e^2}
\end{equation}
where the dimensionless electric and 
magnetic field strength are given by
\begin{equation}
\label{dimensionlesseb}
e=\frac{E}{E_c}, \hspace{1cm} 
b=\frac{B}{E_c}.
\end{equation}

To get an NCYM, we need to take 
$\alpha^{\prime}\to 0$ to decouple massive
open string excitations,  while keeping
the open string moduli $G_{\mu\nu}$,
$\theta^{\mu\nu}$ and $G_s$ finite.
Inspection of Eqs. (\ref{GmunuBperp})
and (\ref{thetaperp}) shows that the following
conditions provide the only possible 
solution for the NCYM limit:
\begin{enumerate}
\item $|e|<1$;
\item $B=bE_c=1/\theta$ finite;
\item $g_0=1$, $g_1=g_2=g \sim 
(\alpha^{\prime})^2$, so that formally 
$E_c$ is a finite parameter; for later 
convenience, to normalize open string 
metric to $G_{11}=G_{22}=1$,
one may take 
$g=(2\pi \alpha^{\prime} B)^2$;
\item $g_{s}\sim\alpha^{\prime}$ to keep $G_s$ finite.
\end{enumerate}

This solution is unique up to finite separate 
rescaling for $g_0$, $g_1$ and $g_2$. It is 
easy to verify that in this limit 
\begin{equation}
\label{thetancym}
\theta^{0i}=0, \qquad \theta^{12}=-\theta.
\end{equation}
Therefore the resulting field theory has only 
space-space noncommutativity. Though $E$
or $e$ does not affect the noncommutativity
parameters $\theta^{\mu\nu}$, it does make
the open string metric $G_{\mu\nu}$
non-diagonal, i.e. it makes the $x_0-$ and
$x_2-$ axes oblique with respect to open 
string metric. The appearance of the 
off-diagonal $G_{02}$ is not surprising: the 
open string endpoint, behaving like a charge, 
acquires a drift velocity in the $x_2$-direction 
in the present cross-field background with 
$E_1=E$ and $B_3=-B$.  

In this way, we see that the scaling limit of  
NCYM is incompatible with space-time 
noncommutativity. This is just right, since 
field theory with space-time noncommutativity
is potentially non-unitary \cite{toum,gomis}. 
A similar analysis can be done for the case with
${\bf E} \parallel {\bf B}$, again resulting 
in an NCYM with vanishing $\theta_{0i}$.
 
\section{Decoupling limit of NCOS}

In this section, we present a new decoupling 
limit of NCOS to demonstrate the interplay 
between the electric and magnetic 
components of the background.

\subsection{The E$\perp$B case}

To achieve this goal, we take in the 
closed string metric Eq. (\ref{metric1})
$g_0=g_1=g$, this leads to corresponding open string
moduli by using Eqs. (\ref{critical})-(\ref{effectiveGper}).

\begin{equation}
\label{aniGmunu}
G_{\mu\nu}=\left(\begin{array}{ccc}
-g(1-e^2)&0&-geb \\
0&g(1-e^2)+\frac{g^2b^2}{g_2}&0
\\-geb&0&g_2+gb^2
\end{array}\right),
\end{equation}
\begin{equation}
\label{anithetaperp}
\theta^{\mu\nu}=
\frac{2\pi\alpha^{\prime}}{g_2g(1-e^2)+g^2b^2}
\left(\begin{array}
{ccc}0&g_2e&0 
\\-g_2e&0&-gb\\0&gb&0
\end{array}\right).
\end{equation}
\begin{equation}
\label{effectiveGperp}
G_s=g_s\sqrt{1-e^2+\frac{gb^2}{g_2}}
\end{equation}

In taking the decoupling limit for NCOS,  
$\alpha^{\prime}$ is kept fixed, while 
$G_{\mu\nu}$ and $\theta^{\mu\nu}$ have 
to have a finite limit. To achieve this goal, we 
introduce the following scaling limit
\begin{enumerate}
\item $e\to 1$, with $g(1-e^2)=
\frac{2\pi\alpha^{\prime}}{\theta_0}$ 
finite;
\item $g_2$ is finite; for convenience, we 
take $g_2=1$;
\item  $b\to 0$, with $gb=
\frac{2\pi\alpha^{\prime}}{\theta_{1}}$ finite.
\end{enumerate}

With this scaling limit, we get the moduli of the 
resulting NCOS as follows: the metric
\begin{equation}
\label{ncosmetric}
G_{\mu\nu}=\left(\begin{array}{ccc}
-\frac{2\pi\alpha^{\prime}}
{\theta_0}&0&
-\frac{2\pi\alpha^{\prime}}{\theta_1} \\
0&\frac{2\pi\alpha^{\prime}}{\theta_0}+
\biggl(\frac{2\pi\alpha^{\prime}}{\theta_1}
\biggr)^2&0\\
-\frac{2\pi\alpha^{\prime}}{\theta_1}&0&1
\end{array}\right),
\end{equation} 
and the noncommutativity matrix 
\begin{equation}
\label{noncomm}
\theta^{\mu\nu}=\frac{2\pi
\alpha^{\prime}}{\frac{2\pi\alpha^{\prime}}{\theta_0}
+\biggl(\frac{2\pi\alpha^{\prime}}{\theta_1}\biggr)^2}
\left(\begin{array}{ccc}
0&1&0 \\
-1&0&-\frac{2\pi\alpha^{\prime}}{\theta_1}\\
0&\frac{2\pi\alpha^{\prime}}{\theta_1}&0
\end{array}\right).
\end{equation}

This scaling limit is striking in that it results 
in NCOS with both space-time and space-space 
noncommutativity. Note that this is different 
from the NCYM limit, where space-time 
noncommutativity can not result from the 
decoupling limit. 
Again, the appearance of nonzero off-diagonal 
elements $G_{02}=G_{20}$ in the NCOS 
metric is very natural: The end points of the 
open string behave like opposite charges 
which, in a cross-field with $E_1$ and $B_3$, 
acquire a drift velocity in the $x_2$-direction 
independent of the sign of charges. (This is 
nothing but the classical picture of the Hall 
effect in condensed matter physics.) 

In the above scaling limit, the open string 
coupling $G_s$ vanishes. To have an interacting 
theory, we can follow Ref. \cite{seiberg} 
to consider $N$ coincident $D$-branes, so that 
the effective open string coupling is
\begin{equation}
\label{effectivecoupling}
G_{eff}=NG_s=Ng_s\sqrt{1-e^2}.
\end{equation}
In the large $N$ limit, if we scale $N$ as
\begin{equation}
\label{largeN}
N\sim\frac{1}{\sqrt{1-e^2}},
\end{equation}
we can keep the effective open string coupling 
$G_{eff}$ finite.

In passing, we emphasize that the decoupling 
conditions $(1)$ and $(3)$ imply that the 
ratio between the electric and magnetic field 
strength is greater than $1$. In other words, 
in our decoupling scheme, the magnetic field 
is held to a finite value. (In fact, the parameter 
$\theta_1$ is just $1/B$). One may wonder 
what will be the NCOS scaling limit if
one assumes $|B|>|E|$. The answer is that 
in this case, we do not have a consistent NCOS 
limit; rather we should take the NCYM limit,
just as we have discussed  in last section.

\subsection{The E$\parallel$B case}

In this subsection, we study the other special 
case where the electric field is parallel to 
the magnetic field. The motivation is to 
show once more that the magnetic effects 
can survive the scaling limit for NCOS, 
resulting in space-space noncommutativity. 
To do so, we choose the closed string metric 
$g_{\mu\nu}$ and anti-symmetric tensor field 
$B_{\mu\nu}$ as
\begin{equation}
\label{gmunupara}
g_{\mu\nu}=\left(\begin{array}{cccc}
-g&0&0&0 \\
0&g&0&0\\
0&0&1&0\\
0&0&0&1
\end{array}\right),
\end{equation}
\begin{equation}
\label{bmunupara}
B_{\mu\nu}=\left(\begin{array}{cccc}
0&E&0&0 \\
-E&0&0&0\\
0&0&0&B\\
0&0&-B&0
\end{array}\right).
\end{equation}

Again, by using Seiberg-Witten relations 
Eqs. (\ref{Gmunu})-(\ref{coup}), we get the 
open string moduli
\begin{equation}
\label{Gmunupara}
G_{\mu\nu}=\left(\begin{array}{cccc}
-g(1-e^2)&0&0&0 \\
0&g(1-e^2)&0&0\\
0&0&1+g^2b^2&0\\
0&0&0&1+g^2b^2
\end{array}\right),
\end{equation}
\begin{equation}
\label{inverse-Gmunupara}
G^{\mu\nu}=\left(\begin{array}{cccc}
-\frac{1}{g(1-e^2)}&0&0&0 \\
0&\frac{1}{g(1-e^2)}&0&0\\
0&0&\frac{1}{1+g^2b^2}&
0\\0&0&0&\frac{1}{1+g^2b^2}
\end{array}\right),
\end{equation}
\begin{equation}
\label{thetapara}
\theta^{\mu\nu}=2\pi\alpha^{\prime}\left(\begin{array}{cccc}
0&-\frac{e}{g(1-e^2)}&0&0 \\
\frac{e}{g(1-e^2)}&0&0&0\\
0&0&0&-\frac{gb}{1+g^2b^2}\\
0&0&\frac{gb}{1+g^2b^2}&0
\end{array}\right),
\end{equation}
\begin{equation}
\label{effectiveGs}
G_{s}=g_s \sqrt{1-e^2} \sqrt{1+g^2b^2}.
\end{equation}
Here we adopted the same conventions for 
$E_c$, $e$, and $b$ as in the previous 
section. From the above open string moduli, 
we see that the same decoupling limit as
that in  ${\bf E}\perp {\bf B}$ case 
can be applied. We also get 
the NCOS with both space-time and space-space 
noncommutativity. In contrast to the 
${\bf E}\perp {\bf B}$ case, the effects of 
the magnetic field is to increase the effective 
open string coupling constant $G_s$ by 
a factor $\sqrt{1+g^2b^2}$ after we take 
the large $N$ limit, without inducing 
a drift motion in other directions.

The most general configuration of $B_{\mu\nu}$ 
can be considered as a superposition of the two 
cases we have discussed, with  ${\bf E}\perp 
{\bf B}$ and ${\bf E}\parallel {\bf B}$ 
respectively. So we conclude that in general,
by decoupling procedure, we can obtain 
NCYM with only space-space noncommutativity, 
or NCOS with both space-time and space-space 
noncommutativity. 

\section{V-duality of NCOS}
 
In the previous section, in the case with 
${\bf E}\perp {\bf B}$, we have managed to get 
a decoupling limit that leads to NCOS with both 
space-space and space-time noncommutativity,
provided that $|{\bf E}|$ is greater than $|{\bf B}|$. 
In electrodynamics it is known that in this case
by a Lorentz boost one can go to a favorable 
inertial frame in which the electromagnetic 
background  becomes purely electric. If we start 
with this frame, the decoupling limit will give us
an NCOS with only space-time noncommutativity.
Before the decoupling limit, our string theory is 
known to have Lorentz symmetry, which allows 
us to transform the gauge field background on the 
D-brane world volume without changing the physics.
So the above argument implies that the NCOS 
theory with both space-space and space-time 
noncommutativity that we obtained in the previous 
section for the case with ${\bf E}\perp {\bf B}$
and $|{\bf E}| > |{\bf B}|$ should be equivalent to
an NCOS with only space-time noncommutativity.
More generally, this argument suggests that NCOS 
theories resulting from electromagnetic backgrounds 
on the D-brane that are related by Lorentz boosts 
should be equivalent to each other. This is a duality
among NCOS with different open string moduli, and it 
is related to Lorentz boosts depending on the relative
{\em Velocity} of the inertial frames. 
We call it {\em V-duality}, so that alphabetically it 
follows the S, T, U dualities we have had already.  

An immediate question is: how V-duality acts on
the open string moduli of NCOS? Now let us try
to determine the orbit of the V-duality action in 
the moduli space of NCOS that we obtained in the 
last section. Let us start with two inertial frames 
$K$ and $K^{\prime}$ on the world volume of 
$D3$-branes, with $K'$ moving relative to $K$
in $x_2$-direction with velocity $v$. Suppose 
the anti-symmetric tensor field $B_{\mu\nu}$ in 
$K$ is purely electric:
\begin{equation}
\label{bmunu}
B_{\mu\nu}=\left(\begin{array}{cccc}
0&E&0&0 \\
-E&0&0&0\\0&0&0&0\\
0&0&0&0
\end{array}\right).
\end{equation}
Then the corresponding $B_{\mu\nu}^{\prime}$
in $K'$ has the form
\begin{equation}
\label{bmunuprime}
B_{\mu\nu}^{\prime}=
\left(\begin{array}{cccc}
0&E^{\prime}&0&0 \\
-E^{\prime}&0&-B^{\prime}&0\\
0&B^{\prime}&0&0\\
0&0&0&0
\end{array}\right).
\end{equation}

To relate $E'$ and $B'$ with $E$, we need to
know the transformation between $K$ and $K'$.
Note that we have taken the metric in both
$K$ and $K'$ to be 
\begin{equation}
\label{ds2}
ds^2=-g(dx_0^2-dx_1^2)+(dx_2^2+dx_3^2).
\end{equation}
To make this metric invariant, the transformation 
should be the following "adapted" Lorentz one:
 \begin{eqnarray}
\label{lorents}
x_2^{\prime}&=&\gamma(x_2-v\sqrt{g}x_0),\\
x_0^{\prime}&=&\gamma(
x_0-\frac{v}{\sqrt{g}}x_2),
\end{eqnarray}
where $\gamma=1/\sqrt{1-v^2}$. It is easy 
to check that the transformed metric is
\begin{equation}
\label{gprimemunu}
-g^{\prime}_{00}=g^{\prime}_{11}=g.
\end{equation}
Using the invariance of the 2-form  
$F=\frac{1}{2}B_{\mu\nu}dx^{\mu}
\wedge dx^{\nu}$, we get the transformed
$B_{\mu\nu}^{\prime}$ in $K'$ as
\begin{eqnarray}
\label{EBprime}
E^{\prime}=\gamma E, \hspace{1cm} 
B^{\prime}= \frac{\gamma v}{\sqrt{g}}E.
\end{eqnarray}

Note that in both $K$ and $K^{\prime}$, 
the definition of $E_c=g/2\pi\alpha^{\prime}$ 
is {\em the same}. Thus we have the 
transformation law for dimensionless electric 
and magnetic fields: 
\begin{eqnarray}
\label{transofdimensionless}
e^{\prime}\equiv\frac{E^{\prime}}{E_c}
=\gamma e, \hspace{1cm}
b^{\prime}\equiv\frac{B^{\prime}}{E_c}
= \frac{\gamma v}{\sqrt{g}}e.
\end{eqnarray}

To study the $V$-duality of NCOS,  now we 
need to take the decoupling limit for NCOS 
in both frame $K$ and $K^{\prime}$. In frame $K$, 
the decoupling limit dictates
\begin{equation}
\label{kdecouple}
g(1-e^2)=\frac{2\pi\alpha^{\prime}}{\theta_0},
\end{equation}
Correspondingly, in frame $K^{\prime}$ we have
\begin{equation}
\label{kprime}
g(1-e^{\prime 2})=\frac{
2\pi\alpha^{\prime}}{\theta^{\prime}_{0}},
\end{equation}
\begin{equation}
\label{gbprime}
g b^{\prime}=\frac{
2\pi\alpha^{\prime}}{\theta^{\prime}_{1}}.
\end{equation}
Thus we can establish the following relation by using 
Eq. (\ref{gbprime}) 
\begin{equation}
\label{equality}
\gamma v \sqrt{g-\frac{2\pi\alpha^{\prime}}{\theta_0}}
= \frac{2\pi\alpha^{\prime}}{\theta^{\prime}_1}.
\end{equation}
The decoupling limit is the one in which
\begin{eqnarray}
\label{decoupling}
e\to 1, \hspace{1cm} g\to\infty,
\end{eqnarray}
So taking the decoupling limit reduces
Eq. (\ref{equality}) to
\begin{equation}
\label{reduced}
v \sqrt{g}=\frac{2\pi
\alpha^{\prime}}{\theta_1^{\prime}},
\end{equation}
Therefore, we conclude that the boost 
velocity $v \to 0$. 

On the other hand, the boost transformation 
 Eq. (\ref{kprime}) leads to
\begin{eqnarray}
\frac{2\pi\alpha^{\prime}
}{\theta_{0}^{\prime}}&=&g(1-e^{\prime 2}) 
\\ \nonumber
&=&g(1-e^2)
+ge^2(1-\gamma^2) \\ \nonumber
&=&g(1-e^2)-\frac{gv^2 e^2}{1-v^2} \\
\nonumber
&\to& \frac{2\pi\alpha^{\prime}}{\theta_0}
-\biggl(\frac{2\pi\alpha^{\prime}}{\theta^{\prime}_1
}\biggr)^2,
\end{eqnarray}
where the arrow $''\to''$ means taking the decoupling 
limit.  Thus, we have proved the $V$-duality 
action for NCOS
\begin{equation}
\label{invrelation}
 \frac{2\pi\alpha^{\prime}}{\theta_0}=
\frac{2\pi\alpha^{\prime}}{\theta_{0}^{\prime}}
+ \biggl(\frac{2\pi\alpha^{\prime}}{\theta^{\prime}_1}
\biggr)^2.
\end{equation}
More generally, with other noncommutativity 
parameters vanishing, the following gives us
an invariant under $V$-duality action:
\begin{equation}
\label{invariant}
\frac{2\pi\alpha^{\prime}}{\theta_{0}}
+ \biggl(\frac{2\pi\alpha^{\prime}}{\theta_1}\biggr)^2
={\rm invariant}.
\end{equation}
This invariance gives us some {\em orbits} 
for $V$-duality. The displacements on an
orbit are determined by the action of group 
elements. This invariance can be viewed as 
a descendant of the Lorentz invariance with 
a boost parameter $v \to 0$, and this is the 
signal of the Galilean group. Therefore, 
we suggest  that the $V$-duality should be 
characterized by a Galilean group or its 
deformation. The 
invariant (\ref{invariant}) is for one of its 
abelian subgroup. 

\section{Holography in NCOS}

Previously in Ref. \cite{LiWu} a 
holographic correspondence between NCYM and 
its supergravity dual was suggested. Namely 
the radial profile of the on-shell close string 
moduli (string-frame metric, NSNS  $B$-tensor 
and dilaton) in the supergravity dual of an 
NCYM can be easily derived through the 
Seiberg-Witten relations 
 \cite{sei-witten} between close 
string moduli and open string moduli, 
provided a simple ansatz for the running 
string tension as the function of the 
energy scale is assumed. In this section, 
we generalize this link between holography 
and noncommutativity to NCOS. 

For convenience of making a contrast between 
NCYM and NCOS, we first briefly recall the case of 
NCYM. Suppose only $B_{23}\neq 0$ on a stack 
of D3-branes. 
The central suggestion made in  Ref. 
 \cite{LiWu} is that in the supergravity 
dual {\it the UV limit (from the NCYM perspective) 
$u\to \infty$ is identified with the NCYM 
"scaling limit" or "decoupling limit"} in Ref. 
 \cite{sei-witten}. In this limit, 
$\alpha'$ should approach zero, as in the AdS/CFT 
correspondence \cite{Mald}. To implement 
this, the overall factor $R^2u^2$, appeared 
in the 4d 
geometry along D3-branes, 
is interpreted as a running string tension 
\begin{equation}
\label{teff}
\alpha'_{run}={1\over R^2u^2},
\end{equation}
which obviously runs to zero in the UV limit.
Note that the manner it approaches zero compared 
to $g_{22}$ and $g_{33}$ agrees with the NCYM 
scaling limit taken in Ref. 
 \cite{sei-witten}. 
The holographic correspondence suggested in 
Ref. \cite{LiWu} is that the {\it 
radial profiles} of the on-shell NSNS fields 
in the gravity dual should {\it satisfy} 
the Seiberg-Witten relations Eqs. (\ref{Gmunu}), 
(\ref{theta}) and (\ref{coup}), with $\alpha'$ 
being replaced by the {\it running} 
$\alpha'_{run}$ given by Eq. (\ref{teff}) and 
with {\it constant (unrenormalized)} open 
string moduli. 

In Ref. \cite{LiWu},  the same holographic 
correspondence was shown to hold for all cases in 
which decoupling leads to an NCYM with {\it space-space} 
noncommutativity and with gravity dual known. 
These include high dimensional D$p$-branes in a 
magnetic background \footnote{When $p\neq 3$, the 
open string (or NCYM) coupling constant is 
no longer $u$-independent: $G_s^2=g^2u^{(7-p)(p-3)/2}$. 
But this just means that the open string coupling runs 
in the same way as in the case when there is no B field, 
in agreement with the result of Ref. \cite{bs} 
in the large-N limit.} and Euclidean D3-branes in a 
self-dual $B$-background. In the following, we would 
like to examine whether a similar holographic 
correspondence holds as well between NCOS (with 
{\it space-time} noncommutativity) and its gravity dual, 
despite that the NCOS limit is very different from the 
NCYM limit. 
   
Let us consider the case with only $B_{01}\neq 0$, In 
this case, because of the existence of a critical 
electric field on the D3-branes, to decouple the 
closed strings, one can no longer take $\alpha'\to 0$. 
Instead, $\alpha'$ is fixed, leading to an NCOS. 
Certainly, the above ansatz Eq. (\ref{teff}) for the 
running string tension $\alpha'_{run}$ should {\it no
longer} hold. We will see that indeed an appropriate 
modification of the ansatz exists, so that the above 
holographic correspondence remains to hold for NCOS.  

The supergravity dual (with Lorentz signature) 
with only $B_{01}$ nonvanishing was given in 
Ref. \cite{gop}:
\begin{eqnarray}
\label{nsf1}
ds^2_{str}&=& H(u)^{-1/2}\left[
{u^4\over R^4}H(u)(-dt^2+dx_1^2)+(dx_2^2+dx_3^2) 
+ H(u)(du^2+u^2 d\Omega_5^2)\right],\\
B_{01}&=&{1\over 2\pi}{u^4\over R^4}, \\
e^{2\phi}&=&g^2{u^4\over R^4} H(u),
\end{eqnarray}
where we have $\alpha'=1$, and $R=4\pi g N$, 
$H(u)\equiv 1+R^4/u^4$. (Again, we omit the 
RR fields.) Recall that in the previous case 
with $B_{23}\neq 0$, the close string 
metric $g_{ij}$ (with $i,j=2,3$) shrinks to zero 
in the UV limit $u\to \infty$. In contrast, in the 
present case the close string metric $g_{\mu\nu}$ 
(with $\mu,\nu=0,1$) goes to infinity in the UV 
limit, being consistent with the NCOS limit
  \cite{seiberg,gop}. So in the spirit of Ref. 
 \cite{LiWu}, we again identify the UV limit 
$u\to \infty$ with the NCOS ``scaling limit'' 
or ``decoupling limit'', assuming the running
string tension of the form
\begin{equation}   
\label{teff1}
\alpha'_{run}= H(u)^{1/2}\equiv 
\left(1+{R^4\over u^4}\right)^{1/2},
\end{equation} 
which is nothing but the inverse of the overall 
factor in front of the bracket\footnote{Inside
the bracket the transverse metric $g_{ij}$ for 
$i,j=2,3$ is taken to be $\delta_{ij}$.} in the 
close string metric in the gravity dual 
 (\ref{nsf1}), in accordance with the same 
prescription as before for Eq. (\ref{teff}).

Now we want to show that the close string moduli 
in Eq. (\ref{nsf1}) can be derived from the 
Seiberg-Witten relations (\ref{Gmunu}) and
(\ref{coup}), in which $\alpha'$ {\it is replaced 
by a running one} given by Eq. (\ref{teff1}).
Introduce the ansatz 
\begin{equation}
\label{ans1}
g_{\mu\nu}=f(u)\eta_{\mu\nu},\qquad 
2\pi B_{\mu\nu}=h(u)\epsilon_{\mu\nu},
\end{equation}
for $\mu, \nu=0,1$, due to the boost symmetry in 
the $(x_0,x_1)$ plane, and assuming constant 
open string moduli:
\begin{equation}
\label{open1}
G_{\mu\nu}=\eta_{\mu\nu}, \qquad 
\theta_{\mu\nu}=2\pi \epsilon_{\mu\nu}. 
\end{equation}
Then two equations in Eq. (\ref{Gmunu}) yield
\begin{eqnarray}
\label{fheq1}
1&=&f-{h^2 H\over f},\\
1&=&{hH \over f^2-h^2H};
\end{eqnarray}
Namely,
\begin{equation}
\label{fs2}
f=f^2-{h^2 H},\hspace{1cm}  f=h H.
\end{equation}
Solving these two equations, one obtains 
\begin{eqnarray}
\label{fs3}
f(u)&=& \left(1-H^{-1}\right)^{-1}
={u^4\over R^4}\left(1+{R^4\over u^4}\right),
\nonumber\\
h(u)&=&{u^4\over R^4}.
\end{eqnarray}

Similarly, substituting the above solution into 
the relation  (\ref{coup}) with the identification 
$G_s=g$, one obtains the $u$-dependent closed 
string coupling: 
\begin{equation}
\label{cco2}
g_s(u)=g\left(\det (g+\alpha'_{run}B)\right)^{1/2},
\end{equation}
or
\begin{equation}
\label{cco3}
e^{2\phi}=g^2 \left(1+{u^4\over R^4}\right).
\end{equation}
The results (\ref{fs3}) and (\ref{cco3}) are 
precisely what appeared in the gravity dual 
 (\ref{nsf1}),  which was previously obtained 
as a solution to classical equations of motion 
in IIB supergravity. Note that the closed string 
coupling approaches unity in the UV limit, in 
agreement with the decoupling of closed strings 
for NCOS. 

Certainly this derivation adds more evidence to 
the universality of the link between holography 
and noncommutativity observed in Ref.  
 \cite{LiWu}. Thus, we have seen that
the relations among the closed string moduli 
and the open string moduli contain much more 
than we could have imagined. With appropriate
ansatz for the input $\alpha_{eff}$, they 
determine the closed string dual of both NCYM 
and NCOS! This demonstrates a simple and direct 
connection between holography and noncommutativity,
either of which is believed to play a role in the 
ultimate theoretical structure for quantum gravity.

\acknowledgments

One of us, GHC, thanks the Institute for Theoretical 
Physics, University of California at Santa Barbara, 
for an ITP Graduate Fellowship, and for the warm 
hospitality he receives during his stay. GHC also 
acknowledges stimulating discussions with Ian Low and 
Miao Li, 
while YSW thanks Feng-Li Lin for discussion. This 
research was supported in part by the National 
Science Foundation under Grants No. PHY94-07194(GHC)
 and PHY-9970701(YSW).  
 

\end{document}